\documentclass[letter]{aa}

\usepackage{graphicx}
\usepackage{natbib}
\usepackage{txfonts}

\begin{document} 
   \title{A VLBI study of the wind-wind collision region in the massive multiple HD\,167971}

  \author{J. Sanchez-Bermudez
          \inst{1,2,3}
          \and
          A. Alberdi\inst{4} \and R. Sch\"odel\inst{4} \and
          W. Brandner\inst{2} \and R. Galv\'an-Madrid\inst{5} \and J.~C. Guirado\inst{6,7} \and R. Herrero-Illana\inst{1} \and C.~A. Hummel\inst{8} 
          \and J.~M. Marcaide\inst{6} \and M.A. P\'erez-Torres\inst{4,9} 
          }

   \institute{European Southern Observatory, Alonso de C\'ordova 3107,Vitacura, Santiago de Chile, Chile \\
\and
             Max-Planck-Institut f\"ur Astronomie, K\"{o}nigstuhl 17,
             D-69117 Heidelberg, Germany
             \and
             Instituto de Astronom\'ia, Universidad Nacional
             Aut\'onoma de M\'exico, Apdo. Postal 70264, Ciudad de
             M\'exico 04510, Mexico \\ \email{joelsb@astro.unam.mx}
         \and
             Instituto de Astrof\'isica de Andaluc\'ia (IAA-CSIC), Glorieta de la Astronom\'ia S/N, 18008, Granada, Spain
             \and
Instituto de Radioastronom\'ia y Astrof\'isica (IRyA), UNAM,
Apdo. Postal 72-3 (Xangari), Morelia, Michoac\'an 58089, Mexico
\and
                         Departament dAstronomia i Astrof\'isica, Universitat de
             Valencia, C. Dr. Moliner 50, 46100 Burjassot, Valencia,
             Spain
             \and
Observatori Astron\`omic, Universitat de Val\`encia, Parc Cient\'{\i}fic, C Catedr\'atico Jos\'e Beltr\'an 2, E-46980 Paterna, Val\`encia, Spain
\and
             European Southern Observatory, Karl-Schwarzschild-Stra$\beta$e 2, D-85748 Garching, Germany
\and
Visiting Scientist: Departamento de F\'isica Te\'orica, Facultad de Ciencias, Universidad de Zaragoza, Spain
             }

\date{Received ; accepted }

\titlerunning{HD\,167971 VLBI radio emission}
 
 \abstract
   {Colliding winds in massive binaries are able to accelerate particles up to relativistic speeds as the result of the interaction between the winds of the different stellar components. HD\,167971 exhibits this phenomenology which makes it a strong radio source.}
   {We aim at characterizing the morphology of the radio emission and
     its dependence on the orbital motion, traced independently by
     NIR-interferometry, of the
     spectroscopic binary and the tertiary component that conforms HD\,167971.}
   {We analyze 2006 and 2016 very long baseline interferometric data at C and
     X bands. We
     complement our analysis with a geometrical model of the wind-wind
     collision region, and with an astrometric description of the system. }
   {We confirm that the detected non-thermal radio emission is associated with the  wind-wind collision region of the spectroscopic binary and the tertiary component in
   HD\,167971. The wind-wind collision region changes orientation in
   agreement with the orbital motion of the tertiary around the
   spectroscopic binary. The total intensity also changes between the two observing epochs in
a way inversely proportional to the separation between the SB and T,
with a negative-steep spectral index typical of an optically thin
synchrotron emission possibly steepened by an inverse Compton cooling effect. The wind-wind collision bow-shock shape and
   its position with respect
   to the stars indicates that the wind momentum from the spectroscopic
   binary is stronger than that of the tertiary. Finally, the astrometric solution derived for
   the stellar system and the wind-wind collision region is consistent
   with independent Gaia data.}
   {}


   \maketitle
%

\begin{figure*}[htp]
\centering
\includegraphics[width=14cm]{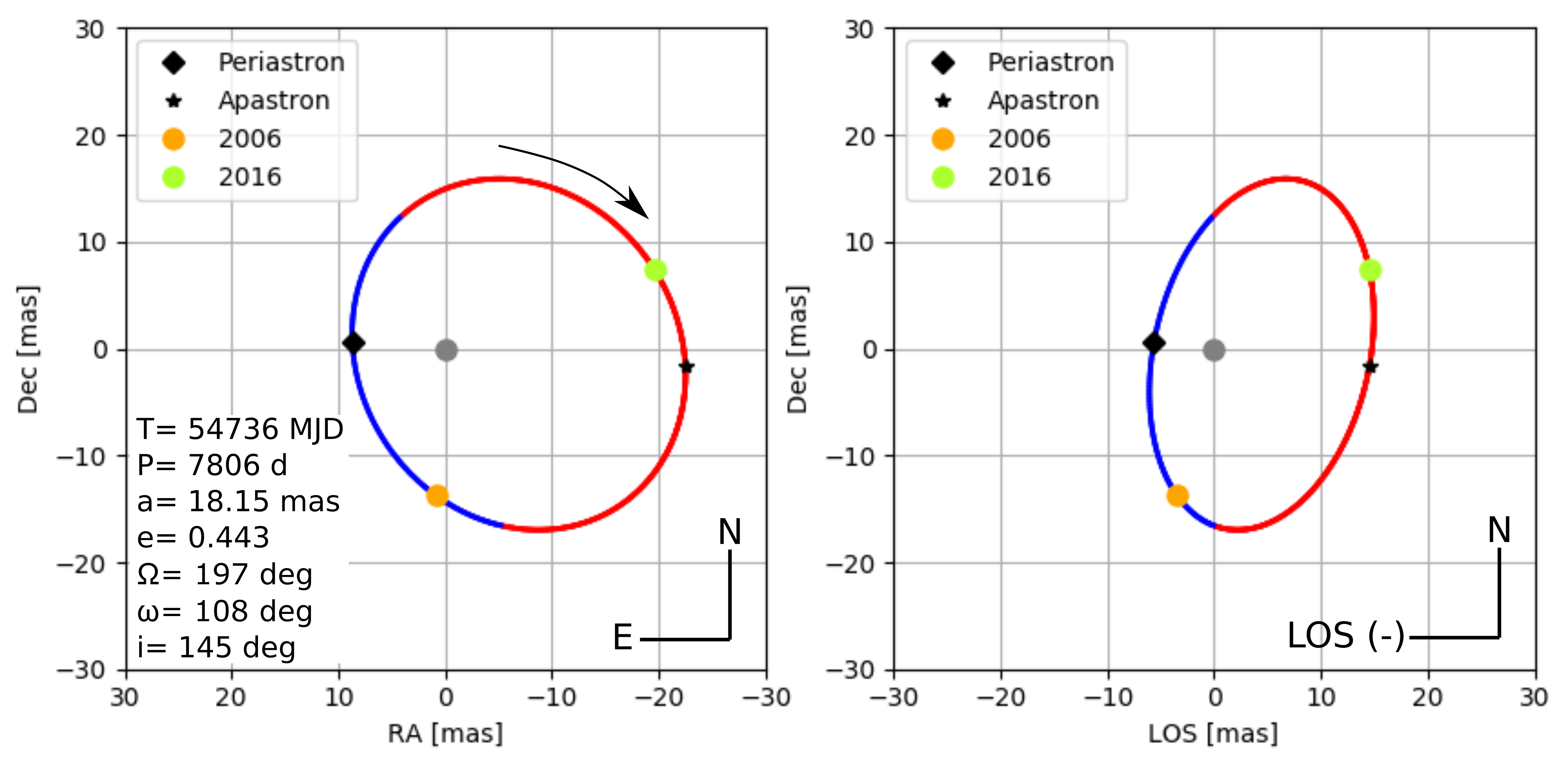}
\caption{\textbf{Left panel:} Orbit of the SB-T system projected in the plane of
  the sky. \textbf{Right panel:} Orbit of the SB-T system in a plane
  orthogonal to the plane of the sky, formed by the
  observer’s line-of-sight and the direction North. The orbital solution was obtained from
  \citet{le_Bouquin_2017} and the mean orbital parameters are labelled on
  the image. The phases of T, at the two VLBI epochs reported in this work, are shown with
  orange and green dots in the orbital solution. The black-arrow shows
  the projected clock-wise motion of the tertiary component. The periastron and
  apastron are also displayed in the panels. The portion of
    the orbit in which T is in front of the SB, in our
  line-of-sight, is displayed in blue; while the portion of the orbit
  in which T is behind the SB is labelled in red.}
\label{fig:orbit}
\end{figure*}

\section{Introduction}

Massive binaries (and higher degree multiple systems) made of O-type
and/or Wolf-Rayet (WR) stars are known to produce stellar wind
collisions and have shown their capability to accelerate particles up
to relativistic speeds. The individual stellar winds
collide, and form shocks that define the wind-wind collision region
(WWCR). Within this region, electrons are accelerated to relativistic
velocities, emitting synchrotron radiation which is detected as
non-thermal radio emission. This emission is characterized by a
negative spectral index\footnote{$S_{\nu}\propto\nu^{\alpha}$; It is assumed that the flux density,
  $S_{\nu}$, is proportional to the frequency of observation, $\nu$, scaling with a power-law defined by the spectral index $\alpha$} and a
high brightness temperature \citep[see e.g., ][]{Eichler_1993,
  Williams_1990, White_1995, Dougherty_1996, Dougherty_2000,
  Dougherty_2003, Pittard_2006, Pittard_2006b, Blomme_2007,
  Montes_2009, Montes_2015}. Most of
the stellar non-thermal radio emitters are multiple systems with at
least one
WR-star. There are only a few known non-thermal systems composed
exclusively by O-stars \citep[see e.g., ][]{Rauw_2004, Benaglia_2007}. The study
of these systems is necessary to understand the properties
of their wind-wind interactions. During the last decade several
individual studies have been done in stellar radio sources like 9 Sgr
\citep{Blomme_2014}, HD\,168112 \citep{DeBecker_2004, Blomme_2005}, or Cyg OB2 No. 9
\citep{vanLoo_2008} among others, confirming that characterizing the
radio emission in terms of the stellar parameters of the sources. 

HD\,167971, at a distance of 1.8 kpc \citep{de_Becker_2012}, is one of the 
brightest synchrotron stellar radio emitters \citep{Bieging_1989}. It is a 
hierarchical triple system \citep{Leitherer_1987}, 
which consists of an spectroscopic binary (SB), with a period 
of $\sim$3.3 days, and a third object (T) moving on a wider orbit,
with a period of $\sim$21.4 years. The spectral types of the stars are
O7.5III (Aa), O9.5III (Ab) and O8I (B)\footnote{The labels Aa, Ab and
  B follows the IAU component designation. However, in this work we
  would refer to the system Aa-Ab as the spectroscopic binary, SB,
  and the component B as the tertiary, T.}. Near-Infrared
interferometric observations \citep[obtained with AMBER, PIONIER and
GRAVITY at the Very Large Telescope Interferometer -VLTI- and reported by
][]{de_Becker_2012, le_Bouquin_2017} demonstrated that the SB and T are gravitationally bound, with a
projected separation that varies from 8 to 15 milliarcseconds
(mas). The angular separation between the components of the SB is of
the order of 0.1 mas, and cannot be spatially resolved directly with
the VLTI nor with radio very long baseline interferometry (VLBI). Therefore, HD\,167971 appears as a binary 
system (SB-T) in the near-infrared interferometric data.  

Figure \ref{fig:orbit} presents the best-fit orbit of
T around the SB reported by \citet{le_Bouquin_2017}. This
solution is in agreement with previous estimates reported by the
light-curve analyses of \citet{Blomme_2007} and
\citet{Ibanoglu_2013}. The VLA and ATCA radio light curves studied by
\citet{Blomme_2007} presented a periodicity consistent with the reported
orbital period of the tertiary, and the emission shown a negative
spectral index, thus confirming its non-thermal nature. These findings
supported the hyphothesis that the non-thermal radio emission is produced 
in the wide orbit, probably, in the adiabatic wind colliding region
between the SB and T. The difference of the light-curve profiles
between 6 cm (5 GHz) and 20 cm (1.5 GHz), suggests that the emitting region is quite extended. 

Additionally to the observed radio emission, \citet{de_Becker_2005}
reported X-Ray (thermal) emission at energies of 2-4 keV that correspond
to a high temperature plasma (2.3 - 4.6 $\times$ 10$^7$ K), which is typical of
pre-shock winds near their terminal velocity. This fact is in
agreement with a wind-wind interaction scenario with strong adiabatic
shocks, and a phase variability scaling with the inverse
of the distance between the SB and T.  These characteristics make HD\,167971 a unique target to test the physics of
wind-wind collisions in O-stars.

Here, we report new HD\,167971 X- (3.5 cm; 8.4 GHz) and C-band (6.0 cm; 5 GHz) observations with the Very Long
Baseline Array (VLBA) obtained in 2016, complemented with archival
VLBA data, at the same frequencies, from 2006. Sect.\,\ref{sec:Observations} presents our observations
and data reduction. In Sect.\,\ref{sec:Analysis} the analysis of the
radio emission and our results are presented. Then, in
Sect.\,\ref{sec:Conclusions} we present the conclusions of our work.


\section{Observations and data reduction \label{sec:Observations}}

We performed VLBA observations of HD\,167971 on August 13th, 2016 at
C- and on August 14th, 2016 at X-band. We used a sustained data recording rate of 2 Gbit/s in two-bit sampling. Each frequency band
was split into eight intermediate frequencies (IFs) of 32 MHz
bandwidth each, for a total synthesized bandwidth of 256 MHz. Each IF
was in turn split into 32 channels of 1.0 MHz bandwidth. The data were
corrected from Earth-orientation and ionospheric effects. The source J1751+0939
was used as bandpass calibrator and, the observations were
phase-referenced to the source J1818-1108. Since this phase calibrator
is not compact, all the fringe-fitting solutions were corrected from its structure. 
 
\begin{figure*}[htp]
\centering
\includegraphics[width=14 cm]{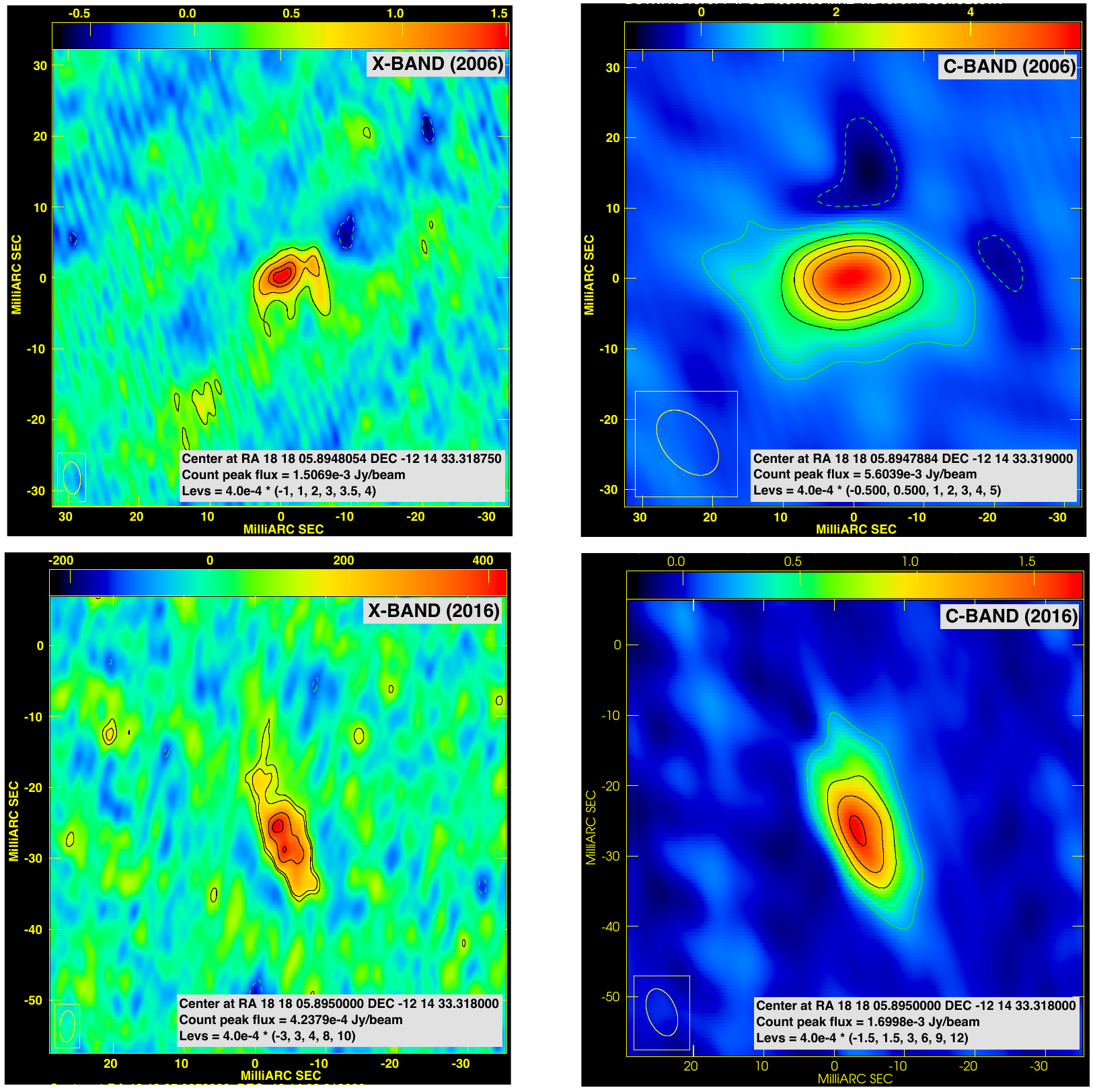}
\caption{Maps of the non-thermal radio emission observed in
  HD\,167971. The radio-bands and epochs are shown in each one of
  the image panels. The synthesized beam, coordinates of the field
  center, peak-flux, and contour levels are also displayed in each one of the panels. }
\label{fig:radio_maps}
\end{figure*}

The data were correlated at the NRAO data processor using an averaging time of 2s.  
We performed standard a-priori gain calibration using the measured
gains and system temperatures of each antenna. This calibration, as
well as the data inspection and editing, were done within the NRAO
Astronomical Image Processing System (AIPS). No self-calibration was
performed on the data since the peaks of emission were too faint. We also used AIPS to produce the images
of both the reference source and target. Additionally, we analyzed archival
VLBA observations performed on August 4th, 2006, half the orbital
period of the system earlier. Then, a total synthesized bandwidth of
32 MHz was used, hence making those observations of lower
sensitivity than those of 2016. The same calibrator source was used, and the
 fringe-fitting solutions were also corrected from the calibrator
 structure. Figure\,\ref{fig:radio_maps} shows the images of both
 epochs that were recovered
by applying natural weighting to the data. Table\,\ref{table:VLBA_obs} displays
the main characteristics of the reconstructed maps.

\begin{table*}
\caption{VLBA Observations}              
\label{table:VLBA_obs}      
\centering                                      
\begin{tabular}{l | c c | c c}          
\hline\hline                        
 & \multicolumn{2}{c|}{2006 epoch}  & \multicolumn{2}{c}{2016 epoch} \\    
\hline                                   
Parameter & X-band & C-band & X-band & C-band\\
\hline
Beam FWHM [mas] & 4.6 $\times$ 2.4  & 10.8 $\times$ 6.9 & 4.38 $\times$ 2.11  & 7.36 $\times$ 3.54
  \\      
Beam PA [deg] & 4.0 & 42.0 & -5.0 & 7.0  \\
rms-noise [$\mu$Jy/beam] & 134.0 & 196.0 & 40.0 & 45.0 \\
Total $S_{\nu}$ [mJy] & 5.324 & 12.314 & 2.817 & 7.316 \\
Peak's $S_{\nu}$ [mJy/beam] & 1.507 & 5.604 & 0.423 & 0.169 \\
\hline                                             
\end{tabular}
\end{table*}

\section{Analysis and Results \label{sec:Analysis}}

\subsection{Radio Maps Analysis}

The reported VLBA observations image the non-thermal radio
emission of the WWCR. Figure \ref{fig:radio_maps} displays the radio
maps, which show that the emission is resolved and consists of an elongated
structure that changes depending on the observing epoch. The observing epochs correspond
to half an orbital period, and show radio structures that have rotated
by almost 90$^{\circ}$. Considering the
orbital solution from \citet{le_Bouquin_2017}, and the radio images, 
we confirm that the long-axis of the radio emission 
--for every frequency and epoch-- is perpendicular to the line that 
connects the SB with T. Our observations, thus, support the hypothesis that the
observed radio emission corresponds to the wind-wind collision region
between the SB and T. The following structural properties are highlighted:

\begin{itemize}
\item The interacting point of the two winds could be characterized
assuming that the WWCR is driven by the wind momenta ratio of the interacting stars
($\beta$=$\dot{\mathrm{M}_{\mathrm{T}}}\nu_{\infty,\,\mathrm{T}} /
\dot{\mathrm{M}_{\mathrm{SB}}}\nu_{\infty,\,\mathrm{SB}}$). In this case, a
bowshock-like structure is formed at the shock front between the two
winds. This bowshock has its tails pointing towards the star with
the smaller wind momentum. In the 2006 epoch, such bow-shock (arc-like)
feature is observed with small tails pointing towards T, implying that
the wind momentum of the SB is larger. 

\item The fact that the bowshock is better defined in the 2006 map is consistent 
with the projected orbital position of T at the predicted orbital phase, 
$\phi$. In the 2006 data\footnote{For the current work, we assume epoch $T_0$ = 2454736.5 JD as reference for the orbital phases } ($\phi_{2006}$=-0.1),
  we are observing the parabolic shock profile with a small
  inclination angle\footnote{Inclination angle, i, is measured
    between the bowshock (rotational) symmetry axis to the plane of
    the sky. This angle increases clock-wise in a plane orthogonal to the plane of the sky in
    direction to the observer's line-of-sight.} ($i$=14.5$^{\circ}$) from the plane of the sky, while in the 2016
  epoch ($\phi_{2016}$=0.37 ; where T is behind the SB in our line-of-sight) we are
  observing the projected apex of the parabolic bowshock pointing towards
  us ($i$=-152.8$^{\circ}$) and, thus, its morphology resembles an elongated Gaussian instead of a
  parabolic arc (see Fig.\,\ref{fig:orbit} for a visual identification
  of the position of T at the two
  analyzed epochs). 

\item In the X-band maps, we identified an additional compact structure to the bowshock. In the 2006 epoch it can be observed as a small, elongated appendix 
to the northwest of the central emission peak. In the 2016 map, a similar appendix 
is also observed to the northeast of 
  the central peak, and the peak itself appears to be shifted from the
  center of the emission. Possible explanations for this include (i) the interaction of
  non-spherical/clumpy winds and, (ii) the possibility of the
  interaction of wind-momenta
  of similar magnitude that could disturb the profile
  from the parabolic bow-shock solution \citep[see Fig.\,2 in][]{Canto_1996}.  

\item We obtained spectral index maps of HD 167971 between 8.4 and 5 GHz 
(see Appendix \ref{sect:app1}). The spectral index corresponds to non-thermal 
emission, with a value of $\alpha \sim$-1.1 quite constant along the whole shocked
region, suggesting that the Fermi acceleration mechanism is
efficient. \citet{Dougherty_2003} show that a synchrotron spectrum,
without losses, goes as $\propto \nu^{-0.5}$. However, HD\,167971 has a
steeper spectral index, which suggest a possible attenuation of the
synchrotron emission due to several physical mechanisms (see below).

\citet{Blomme_2007} determined, based on typical stellar parameters
  for O-stars, that the radio photosphere of optical depth unity at 5 GHz is $\leq 2000
  R_{\odot}$ (at 8.4 GHz, it is smaller), shorter than the minimum linear
  separation between the eclipsing binary and the tertiary component
  of the system ($\sim$ 4000  $R_{\odot}$). Therefore, the observed WWCR should be outside the radio photosphere,
with a negligible free-free absorption, showing the intrinsic optically
thin synchrotron emission at all orbital phases. It is predicted that at
  frequencies lower than 5 GHz (e.g., at 1.4 GHz), the radio photosphere 
  is much more extended. Therefore, the radio emission of the WWCR is
  embedded in the radio photosphere for a large fraction of the orbit, resulting in spectral index and turnover frequency changes of the radio 
  emission along the orbit, which are not detectable at 5/8.4
  GHz. Under these assumptions, attenuation of the synchrotron emission
  at the observed frequencies due to the Razin effect
  \citep{Dougherty_2003} is also negligible. 

  We suspect that the steepness of $\alpha$ is caused by
  inverse Compton cooling, which is more efficient for electrons with
  higher energies. According to the model
  presented by \citet{Pittard_2006b}, inverse Compton decreases
  rapidly the energy of electrons close to the central volume of the
  WWCR, confining only a portion or relatively energetic electrons in
  layers near the shocks, which, in turn, results in a steeper
  spectrum at high frequencies.  

\end{itemize}

\subsection{Bow-shock Model}

Since the projected bow-shock profile in the 2006 epoch does not
  exhibit long tails, despite the small inclination angle from the
  plane of the sky, we suspect that the wind-momenta ratio of the SB
  and T are quite similar. To estimate the separation of the WWCR and $\beta$,
we used the algebraic solution of \citet{Canto_1996} to define the
shock profile of two
interacting winds:

\begin{equation}
R = \mathrm{D}\,\mathrm{sin}(\theta_{\mathrm{SB}})\,\mathrm{csc}(\theta_{\mathrm{SB}}+\theta_{\mathrm{T}})\,
\end{equation}

where, in our system, $R$ corresponds to the bow-shock distance from
T, and $D$ is the separation between the SB and T. The distance from
$T$ to the apex of the bowshock is
known as the stand-off distance $R_0$. The angles $\theta_{\mathrm{T}}$ and
$\theta_{\mathrm{SB}}$ are the different half-opening angles between
the line that connects the SB and T with the profile of the wind-wind
shock front. The half-opening angles depend on $\beta$, as follows:

\begin{equation}
\theta_{\mathrm{SB}}\,\mathrm{cot}(\theta_{\mathrm{SB}}) = 1+\beta\,(\theta_{\mathrm{T}}\,\mathrm{cot}(\theta_{\mathrm{T}})-1)\,
\end{equation}

Since the wind-wind shock profile is more homogeneous in the C-band,
we used the 2006 map to model the bow-shock
profile with the previous geometrical solution through the following method:

\begin{itemize}
\item We mapped the position of the maximum of the source's brightness
  distribution for each column of pixels, perpendicular to the
  bow-shock structure, in the 2006 C-band image. A
  prior rotation according to the orbital position, and image centering
  in a common reference frame were applied to the original map. We restricted
  our analysis to the structure with a brightness larger than 20\% of
  the flux peak. To map the bow-shock profile, a Gaussian was fit to each non-zero element per column of pixels in the image. To estimate the errors in the position of the bow-shock profile, the standard-deviation of the Gaussian fit was used with a weight proportional to the number of pixel elements fitted.
  
  \item Once the profile was estimated, it was compared with the
    projected canonical solution of the wind-wind bowshock obtained
    from a Monte-Carlo Markov-Chain (MCMC) model using \texttt{emcee}
    \citep{Foreman_2013}. Our MCMC model evaluates values of $\beta$
    between 0.3 and 0.9, smaller $\beta$ values would produce a very
    close bowshock with long tails, while higher ones would produce a
    completely open bowshock. The model was run with ten independent chains with 1000 steps each one. For every iteration the distance between T and the shock front was adjusted for both the model and data, according to the following expression:  
  
  \begin{equation}
R_0 = \frac{\beta^{1/2}\mathrm{D}}{1+\beta^{1/2}}\,
\end{equation}

\end{itemize}

\begin{figure}[htp]
\centering
\includegraphics[width=\columnwidth]{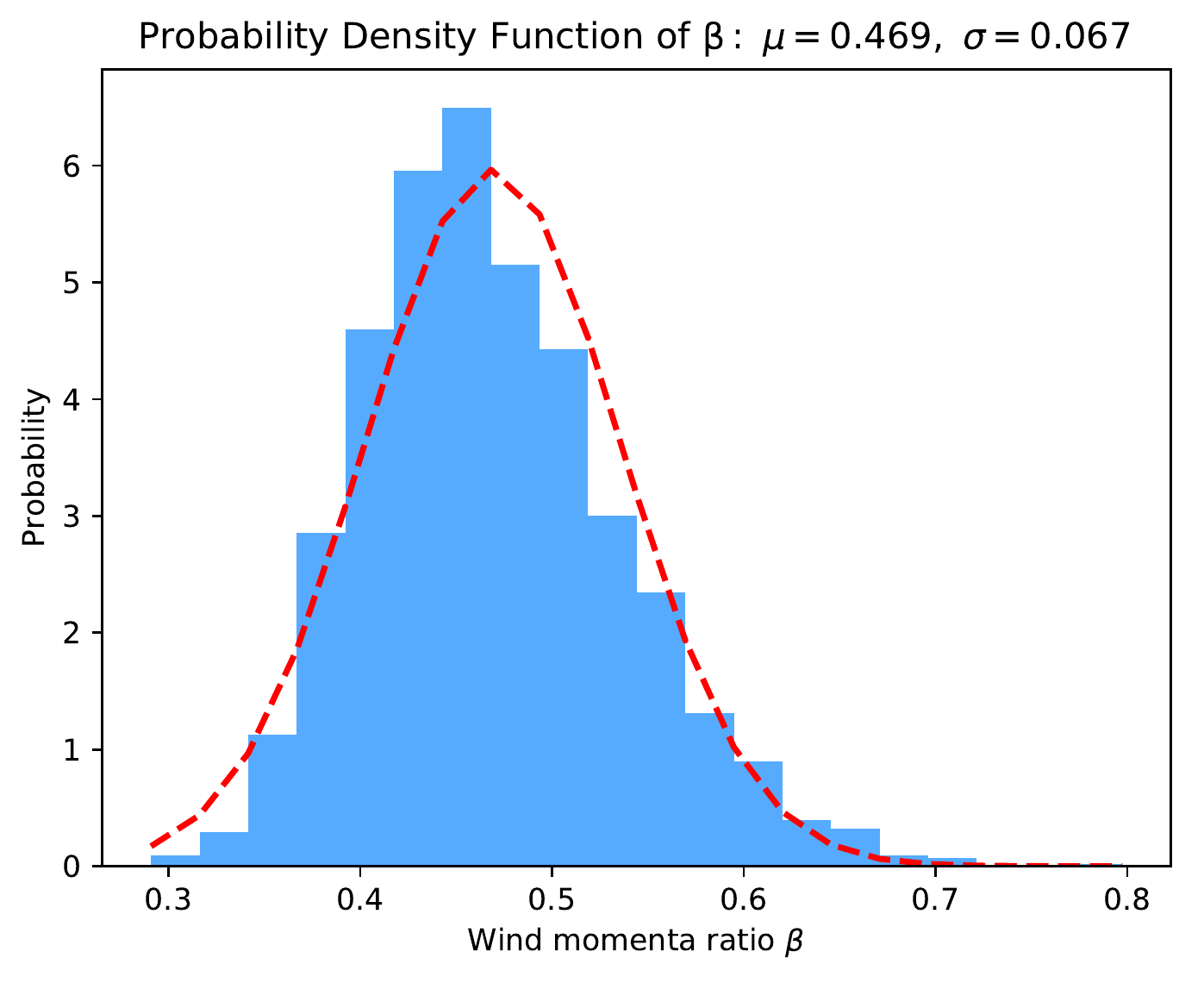}
\caption{Probability density distribution of $\beta$. The values were obtained from the MCMC minimization of the geometrical modelling applied to the wind-wind bow-shock profile of HD\,167971.}
\label{fig:chi2_map}
\end{figure}

Although the stand-off distance is variable in this procedure since
the half-opening angle of the bowshock depends on the wind momenta
ratio, we found the most probable model at a $\beta$ and a stand-off
distance that match better the half-opening angle of the extracted
profile. The probability density function of the MCMC exhibits a clear
  peak at $\beta$=0.48$\pm$0.07 (see Fig\,\ref{fig:chi2_map}), which
  locates the shock front at a projected $R_0$=0.41D$\pm$0.02D
  from T, being D the separation between the SB and T across the
  orbital path. The $\beta$ value corresponds, thus, to a
  half-opening angle $\theta_{\mathrm{T}}$ = 103$^{\circ} \pm$ 3$^{\circ}$ of
  the bow-shock profile.

The used model is a solution for a radiative wind-wind
  collision shock. However, it is quite probable for the shock in HD\,167971 to be
  adiabatic. Therefore, we compared the model solutions for adiabatic
  shocks described in \citet{Gayley_2009} \citep[see also][]{Pittard_2018}. The model for an
  adiabatic shock without mixing requires a  $\beta \sim$ 0.39 (notice
  that this value is within 2$\sigma$ of
  the reported $\beta$ value for the radiative shock) to
  generate a half-opening angle of  $\theta_{\mathrm{T}} \sim$103$^{\circ}$. However, the situation
changes when there is an adiabatic shock with mixing. In this case,
the value of $\beta$ depends on the
ratio of the terminal velocity of the winds, $u$=$\nu_P/\nu_S$. The
bigger the value of $u$ is, smaller values of $\beta$ are required to
keep the required half-oppening angle (see Fig.\,\ref{fig:bs_profiles}). In turn, the
position of the shock relative to the two stellar components would be
modified considerably. It is beyond
the scope of this work to explore in detail these models. However,
they should be taken into account for future studies.

\begin{table*}[htp]
\caption{HD\,167971 astrometry}              
\label{table:VLBA_astrometry}      
\centering                                      
\begin{tabular}{l  c c  c c}          
\hline\hline                        
\multicolumn{5}{l}{\textbf{VLBI coordinates of the emission's peak}}  \\    
Epoch & R.A. &  $\sigma_{\mathrm{RA}}$ & Dec. & $\sigma_{\mathrm{Dec}}$\\
\hline
2006.59& 274.5245618$^{\circ}$  & 0.4 mas & -12.2425887$^{\circ}$ & 0.2 mas\\
2016.61& 274.5245618$^{\circ}$  & 1.9 mas & -12.2425957$^{\circ}$ & 0.9 mas\\
\hline
\hline
\multicolumn{5}{l}{\textbf{SB relative position from the VLBI coordinates}$^\mathrm{a}$}  \\ 
Epoch & $\Delta$R.A. &  $\sigma_{\mathrm{\Delta RA}}$ & $\Delta$Dec. &
                                                                       $\sigma_{\mathrm{\Delta Dec}}$\\
\hline
2006.59 & -0.4 mas  & 0.4 mas & 7.0 mas & 0.3 mas\\
2016.61 & -10.0 mas  & 2.0 mas & -3.8 mas & 1.0 mas\\
\hline
\hline   
\multicolumn{5}{l}{\textbf{Gaia coordinates}$^\mathrm{b}$}  \\ 
Epoch & R.A. &  $\sigma_{\mathrm{RA}}$ & Dec. & $\sigma_{\mathrm{Dec}}$\\
\hline
2016.61 & 274.5245619$^{\circ}$  & 0.2 mas & -12.2425958$^{\circ}$ & 0.2 mas\\
\hline
\end{tabular}
\begin{list}{}{}\footnotesize
\item[$^\mathrm{a}$] The reported errorbars have in consideration the error reported for the position of the binary system with respect to the radio emission derived from our model, and the intrinsic astrometric error of the radio maps.
\item[$^\mathrm{b}$] The reported errorbars include the Gaia
  uncertainty of the 2015.5 position and the uncertainties in the
  proper motions.
\end{list}
\end{table*}

\subsection{Astrometric solution}

\begin{figure}[htp]
\centering
\includegraphics[width=9 cm]{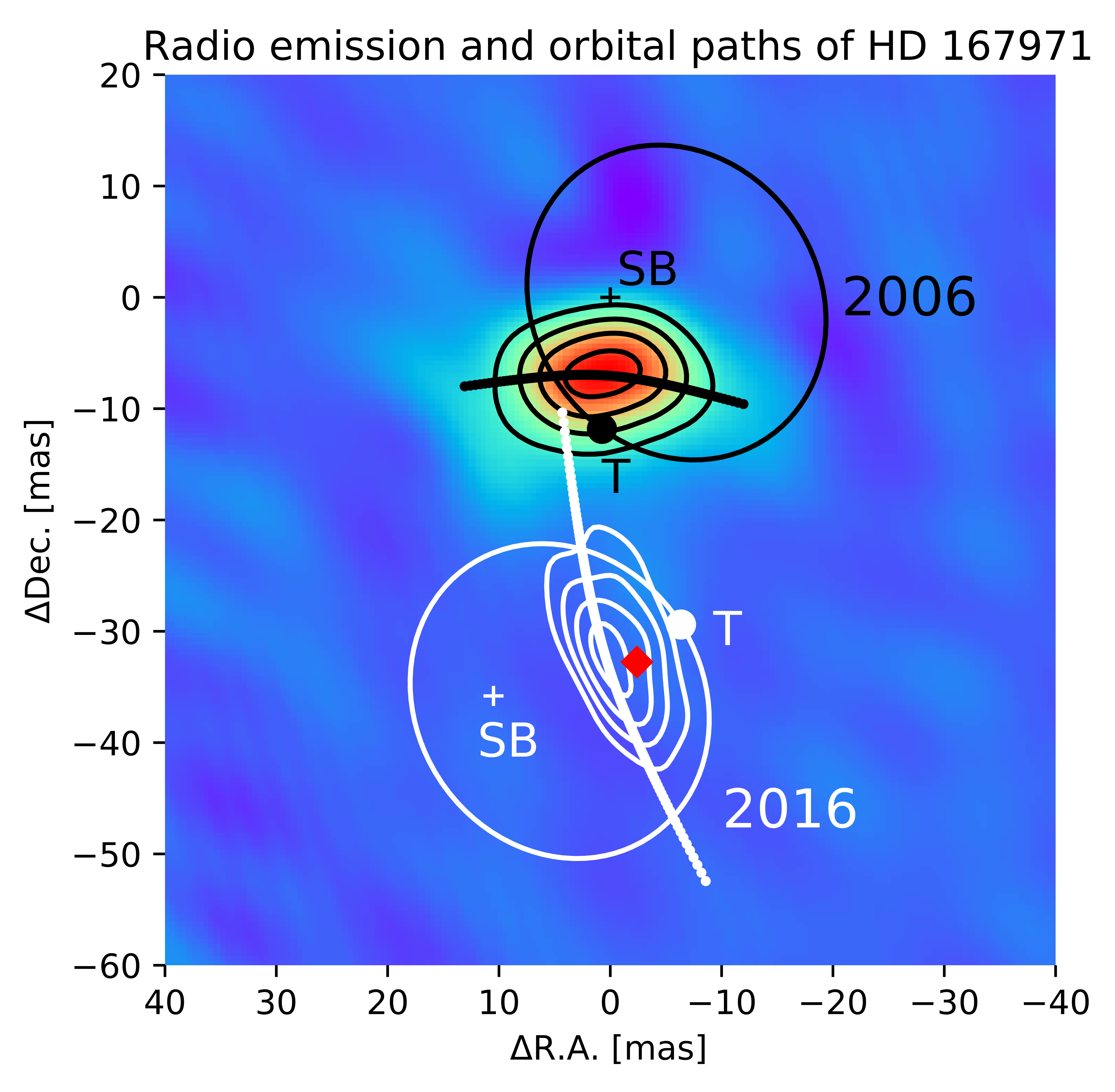}
\caption{Astrometric position of the SB-T system and the C-band radio
  emission of the 2006 and 2016 epochs. The image is centered at
  R.A.= 274.5245616, Dec.= -12.2425868. The different stellar
  components and the radio emission are represented in black for the
  2006 epoch, and in white for the 2016 one. The SB is represented by a cross and T with a dot. The projected orbital path is
  shown with a solid line. The 2006 C-band radio-emission is in
  colour scale contoured in black (representing 30, 50, 70, and 90\%
  of the peak), while the 2016 one is only displayed in white contours. The
  thick-solid lines represent the 2D bow-shock profile from our
  geometrical model. The red diamond shows the Gaia
  astrometric measurement for the system in the 2016 epoch.   }
\label{fig:astrometry}
\end{figure}

Our VLBI radio observations provide us with the absolute astrometric position of
the bowshock and our model allowed us to locate the projected
position of the binary referenced to the radio emission. In
Table\,\ref{table:VLBA_astrometry}, the positions
of the emission's peaks of the radio maps are reported, together with
the relative position of the SB from its respective radio maps (as computed from our
model). We confirmed our 2016 astrometric
solution by comparing the position of the system reported in the Gaia Data Release 2 archive \citep{Gaia1_2016, Gaia2_2016}. The Gaia
2015.5 epoch reported a R.A.= 274.5245621 and Dec.=-12.2425954 (with
uncertainties around 0.1mas) for the position of
HD\,167971. Correcting by the barycentric proper motions
(PM$_{\mathrm{R.A.}}^{\mathrm{sys}}$=-0.54$\pm$0.19 mas yr$^{-1}$;
PM$_{\mathrm{Dec}}^{\mathrm{sys}}$=-1.25$\pm$0.17 mas yr$^{-1}$), we
derived a mean Gaia astrometric position for the 2016.61 radio-epoch
(see Table\,\ref{table:VLBA_astrometry}). These new coordinates set the
Gaia astrometric solution in between the line that connects the SB and T,
confirming the astrometry derived from our VLTI binary fit and VLBA radio emission (see Fig.\,\ref{fig:astrometry}).  

\section{Conclusions \label{sec:Conclusions}}

In this paper we have shown that the non-thermal radio emission
observed in the C and X bands is in agreement with the wind-wind
collision region between the spectroscopic binary
and, the more distant, tertiary component in HD\,167971. 
The morphology of the emission changes in accordance with the 
predicted orbital motion of the components as determined from
NIR interferometric observations,
with the semi-major axis perpendicular to the projected line that
connects the SB and T across the orbital path. The total intensity also 
changes accordingly between the two observing epochs in a way 
inversely proportional to the separation of the stellar
components in the long-period system, with a steep spectral index of
$\alpha \sim$ -1.1. Since the WWCR is outside the radio photosphere of the
SB and T for all the orbital path, free-free absorption and the
Razin effect are excluded as attenuating mechanisms of
the synchrotron emission at the observed frequencies. In contrast
inverse Compton cooling appears to play an important role in the
steepness of $\alpha$. Our model allowed us to locate the
orbital components relative to the bow-shock position, finding an absolute astrometric solution for the system, which, in turn,
was confirmed with an independent Gaia measurement. 

This work presents the first step for understanding the physics of the 
observed synchrotron radio emission of HD\,167971. It has shown the 
importance of VLBI observations to characterize the wind-wind interaction 
region in the system. Future multi-wavelength VLBI radio observations are 
necessary to fully characterize the emission across the different orbital
phases. Monitoring the system would be important to constrain the
stellar parameters of the components and to predict, through radiative
transfer modelling, its evolution. 
Complementary observations with the high-resolution
(R$\sim$4000) mode of GRAVITY/VLTI are also envisioned to characterize
possible infrared shock tracers (like Br$\gamma$) correlated with the observed
WWCR at radio frequencies \citep[see e.g., ][]{Sanchez-Bermudez_2017,
  Sanchez-Bermudez_2018}. HD\,167971 is a unique target that can
bring us new insights about the synchrotron emission in multiple O-star
systems. Understanding the properties of the stars, could help us to
better characterize more distant targets in high-mass star forming regions
\citep[e.g., the stellar radio emitters in W51;][]{Ginsburg_2016},
which in turn would provide important constraints on the radiation and
chemical feedback of massive stars into the inter-stellar medium.

\begin{acknowledgements}
  We thank the anonymous referee for the very useful comments. J.S.B acknowledges support from the ESO Fellowship program.This
      work made use of data from the European Space Agency (ESA) mission
{\it Gaia} (\url{https://www.cosmos.esa.int/gaia}), processed by the {\it Gaia}
Data Processing and Analysis Consortium (DPAC,
\url{https://www.cosmos.esa.int/web/gaia/dpac/consortium}). Funding for the DPAC
has been provided by national institutions, in particular the institutions
participating in the {\it Gaia} Multilateral Agreement. The research
leading to these results has received funding from the European
Research Council under the European Union's Seventh Framework
Programme (FP7/2007-2013) / ERC grant agreement no. [614922]. A.A.,M.P.-T. acknowledge support from the Spanish MINECO through
grant AYA2015-63939-C2-1P, cofunded with FEDER funds. J.C.G., and
J.M.M. were partially supported by the Spanish MINECO project
AYA2015-63939-C2-2-P. R.G.M. and J.S.B. acknowledge support from UNAM-PAPIIT Programme IN104319.
\end{acknowledgements} 


\bibliography{/Users/bluedemon/Documents/Papers/Paper_lib}

\appendix

\section{HD\,167971 spectral index map \label{sect:app1}}

\begin{figure}[htp]
\centering
\includegraphics[width=\columnwidth]{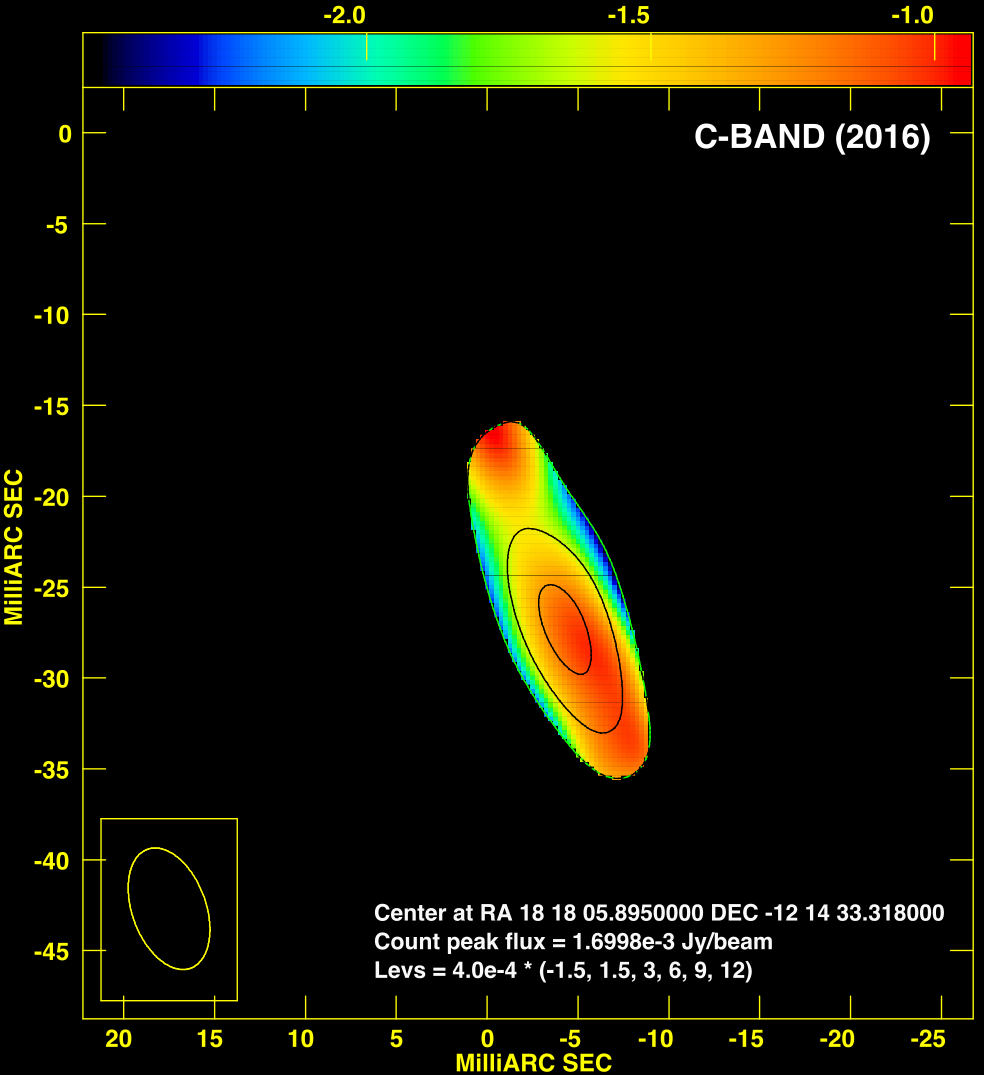}
\caption{ In colours it is represented the spectral index of HD\,167971
  between X-band and C-band of the 2016 epoch. In contours (representing 30, 50, 70, and 90\%
  of the peak) the image displays the structure of
  HD\,167971 observed at the X-band. Notice how the index across the whole
  structure is quite consistent around $\alpha$=-1.1, confirming the non-thermal
  emission of the observed radio emission.
}
\end{figure}

\section{Bowshock profiles \label{sect:bs_prof}}

\begin{figure}[htp]
\centering
\includegraphics[width=\columnwidth]{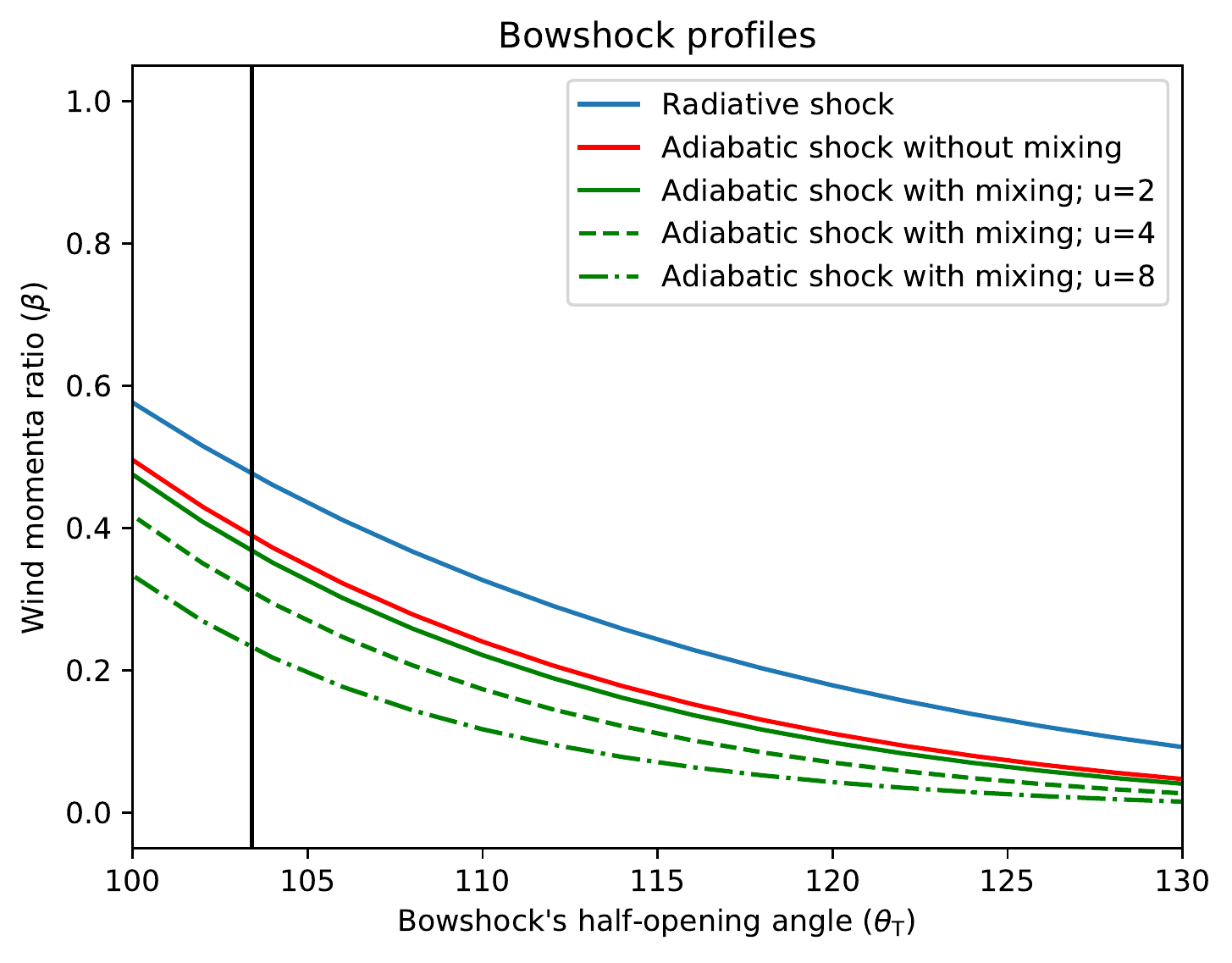}
\caption{ The plot shows the value of $\beta$ as function of the
  half-opening angle ($\theta_{\mathrm{T}}$) of the bowshock profile for different types of
  wind-wind collision shocks according with \citet{Canto_1996}
  (radiative) and \citet{Gayley_2009} (adiabatic). The labels on
  the frame describe each
  one of the different shock types plotted. In the case of the
  adiabatic shock with mixing, three different cases are shown for
  different values of terminal velocity ratio, $u$. The vertical
  black-solid line shows the corresponding $\beta$ values for
  $\theta_{\mathrm{T}}$ = 103.4$^{\circ}$, which is the best-fit value for the
  shock profile obtained from our 2006 C-band image. 
}
\label{fig:bs_profiles}
\end{figure}

\end{document}